\documentclass[final,3p,times,twocolumn]{elsarticle}

\usepackage{amssymb}
\usepackage{graphicx}
\usepackage{subfig}
\usepackage{natbib}
\usepackage{color}

\newcommand*{\pbar}{\ensuremath{\overline{\rm{p}}}}
\newcommand*{\pbHe}{\ensuremath{\overline{\rm{p}}\rm{He}^+}}

\journal{Physics Letters B}

\begin{document}

\begin{frontmatter}



\title{Antiproton magnetic moment determined from the HFS of $\pbHe$}


\author[1]{T.~Pask}
\author[2,3]{D.~Barna}
\author[2]{A.~Dax}
\author[2]{R.~S.~Hayano}
\author[2,4]{M.~Hori}
\author[3,5]{D.~Horv\'ath}
\author[1]{S.~Friedreich}
\author[1]{B.~Juh\'asz}
\author[1]{O.~Massiczek}
\author[2]{N.~Ono}
\author[3,4]{A.~S\'ot\'er}
\author[1]{E.~Widmann}

\address[1]{Stefan Meyer Institute for Subatomic Physics, Austrian Academy of Sciences, Boltzmanngasse 3, A-1090 Vienna, Austria.}
\address[2]{Department of Physics, University of Tokyo, 7-3-1 Hongo, Bunkyo-ku, Tokyo 113-0033, Japan.}
\address[3]{KFKI Research Institute for Particle and Nuclear Physics, H-1525 Budapest, PO Box 49, Hungary.}
\address[4]{Max-Planck-Institut f\"{u}r Quantenoptik, Hans-Kopfermann-Strasse 1, D-85748 Garching, Germany.}
\address[5]{Institute of Nuclear Research of the Hungarian Academy of Sciences, H-4001 Debrecen, PO Box 51, Hungary}

\begin{abstract}
We report a determination of the antiproton magnetic moment, measured in a three-body system, independent of previous experiments.  We present results from a systematic study of the hyperfine (HF) structure of antiprotonic helium where we have achieved a precision more than a factor of 10 better than our first measurement.  A comparison between the experimental results and three-body quantum electrodynamic (QED) calculations leads to a new value for the antiproton magnetic moment $\mu^{\pbar}_{s} \, = \, -2.7862 (83) \mu_{N}$, which agrees with the magnetic moment of the proton within 2.9~$\times$~10$^{-3}$.  
\end{abstract}


\end{frontmatter}


Antiprotonic helium ($\pbHe$) is a three body system consisting of an antiproton, a helium nucleus and an electron ($\pbar$, He$^{++}$, e$^-$)~\cite{Yamazaki:93,Yamazaki:02,Hayano:2007}.  It is formed by stopping antiprotons in a helium medium.  Because of its mass, the $\pbar$ occupies states of principle quantum number $n$~$\sim$~38 with the highest probability, while the e$^-$ remains in the ground state.  The vast proportion of newly formed $\pbHe$ atoms proceed rapidly (within nanoseconds) to an ionised state by Auger excitation of the electron.  The antiproton then annihilates almost instantaneously (within picoseconds) with one of the nucleons in the helium nucleus due to the overlap of their wavefunctions.  However, a small proportion ($\sim$~3\%) occupy circular states $n$~$\sim$~$l$, where $l$ is the total angular momentum quantum number.  Auger decay is then supressed by the large ionisation energy ($\sim~25$~eV) and degeneracy is lifted due to the presence of the electron~\cite{Condo:64}.  These states become relatively long lived (metastable) because the only decay channel available to them is the radiative one, and they cascade from ($n$,~$l$)~$\rightarrow$~($n-1$,~$l-1$) with typical energy level spacings $\sim$~2~eV and lifetimes of $\sim$~1.5~$\mu$s.

Shortly after its discovery, in 1991~\cite{Iwasaki:91}, laser spectroscopy measurements were performed on various levels of the $\pbHe$ cascade~\cite{Morita:94,Torii:99,Hori:01,Hori:03}.  Compared with three-body Quantum Electrodynamic (QED) calculations these measurements were used to determine the antiproton-to-electron mass ratio ($m_{\pbar}/m_{\rm{e}}$).   Over the years, improvements to the experimental system have increased the precision to 2~ppb~\cite{Hori:06}, which is one of the best tests of CPT invariance in the baryon sector.  In 1997 the hyperfine structure was first revealed by laser spectroscopy~\cite{Widmann:97} and, in 2002, measured via a laser-microwave-laser technique~\cite{Widmann:02}.

This paper concludes a systematic study, commenced in 2006~\cite{Pask:2008}, where the error has been reduced by more than a factor of 10 over the first measurement~\cite{Widmann:02} and a new value for the antiproton spin magnetic moment $\mu^{\pbar}_{s}$ has been determined by comparison with QED calculations.

\begin{figure}[t]
\begin{center}
\includegraphics[scale=0.4]{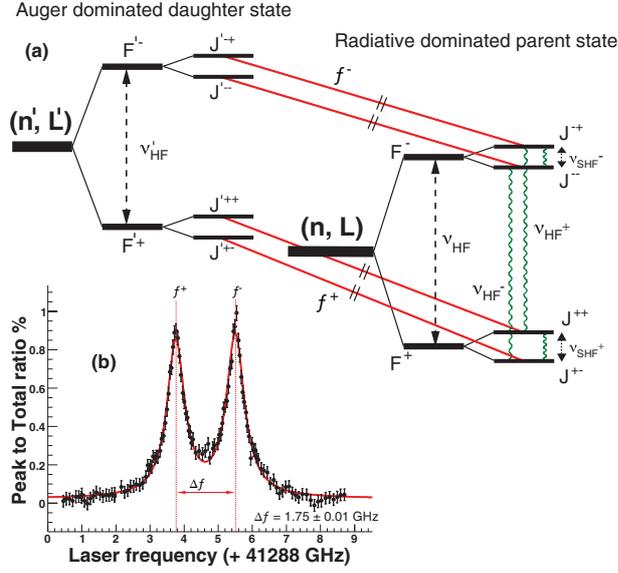}
\end{center}
\caption {(a) Schematic view of the level splitting of $\pbHe$ for the ($n$,~$l$)~$\rightarrow$~($n$~$-$~1,~$l$~+~1) electric dipole transitions.  The laser transitions $f^{+}$ and $f^{-}$, from the parent to daughter states, are indicated by straight lines and the microwave transitions, between the quadruplets of the parent, by wavy ones.  For this experiment $(n,L)$~=~(37,~35) and $(n',L')$~=~(38,~34). (b) Laser resonance profile demonstrating the two sharp peaks and HF laser splitting $\Delta f$~$\equiv$~$f^- \, - \, f^+$.  Although there are four SHF laser transitions only the HF ones can be resolved in this experiment.}
\label{fig:HF_trans}
\end{figure}

The \emph{hyperfine (HF)} splitting~\cite{Yamazaki:02} of $\pbHe$ arises from the coupling of the e$^-$ spin $\vec{S}_{e}$ with the $\pbar$ orbital angular momentum $\vert \vec{L} \vert$~$\sim$~35$\hbar$ and results in a doublet structure of the order $\nu_{\rm{HF}}$~=~10~-~15~GHz.  The interaction between the antiproton spin $\vec{S}_{\pbar}$ with $\vec{S}_{e}$ and $\vec{L}$ causes a \emph{superhyperfine (SHF)} splitting of size $\nu_{\rm{SHF}}$~=~150~-~300~MHz.  A schematic of the energy level structure is presented in Fig.~\ref{fig:HF_trans}a.  

The theoretical framework for the level splitting has been developed by two separate groups~\cite{Bakalov:98,Korobov:01,Yamanaka:01,Kino:03APAC} which all use the same Hamiltonian, first derived by Bakalov and Korobov~\cite{Bakalov:98}, but different variational methods to extract the energy eigen values.

The HF doublet is characterised by the quantum number $\vec{F}$~=~$\vec{L}$~+~$\vec{S}_{e}$ and the SHF quadruplet by $\vec{J}$~=~$\vec{F}$~+~$\vec{S}_{\pbar}$ = $\vec{L}$~+~$\vec{S}_{e}$~+~$\vec{S}_{\pbar}$.  An electron spin flip transition can be induced by an oscillating magnetic field, resulting in two M1 transitions: $\nu^{+}_{\rm{HF}}: \, J^{++} \, = F^{+} \, + \, \frac{1}{2} \,  \longleftrightarrow \, J^{-+} \, = F^{-} \, + \, \frac{1}{2}$ and $\nu^{-}_{\rm{HF}}: \, J^{+-} \, = F^{+} \, - \, \frac{1}{2} \, \longleftrightarrow \, J^{--} \, = F^{-} \, - \, \frac{1}{2}$

Bakalov and Widmann~\cite{Bakalov:07} indicate the sensitivity of certain states on $\mu^{\pbar}_{s}$ and the precision required to improve its value over the most precise measurement~\cite{Kreissl:88}.  Some of these states are not practical due to limitations in laser capability. Others, like the ($n$,~$L$)~=~(39,~35) state, which are within laser capabilities, have a HF laser splitting $\Delta f$~=~0.5 GHz, of the same order as the Doppler broadening $\Delta f_{D}$~=~420~MHz. The previously measured (37,~35) state remains the best candidate for a precision study because there is an easily stimulated laser transition between the ($n$,~$L$)~=~(37,~35) and (38,~34) states with $\Delta f$~=~1.75~GHz.  It is therefore possible to individually resolve the $F^{\pm}$ states, demonstrated in Fig.~\ref{fig:HF_trans}b, since $\Delta f_{D}$~=~320~MHz. The parent state is also relatively highly populated, containing some 0.3\% of the antiprotons stopped in the target~\cite{Hori:02}.

\begin{figure}[t]
\begin{center}
\subfloat[]{
\label{fig:subfig:timing}
\includegraphics[scale=0.5]{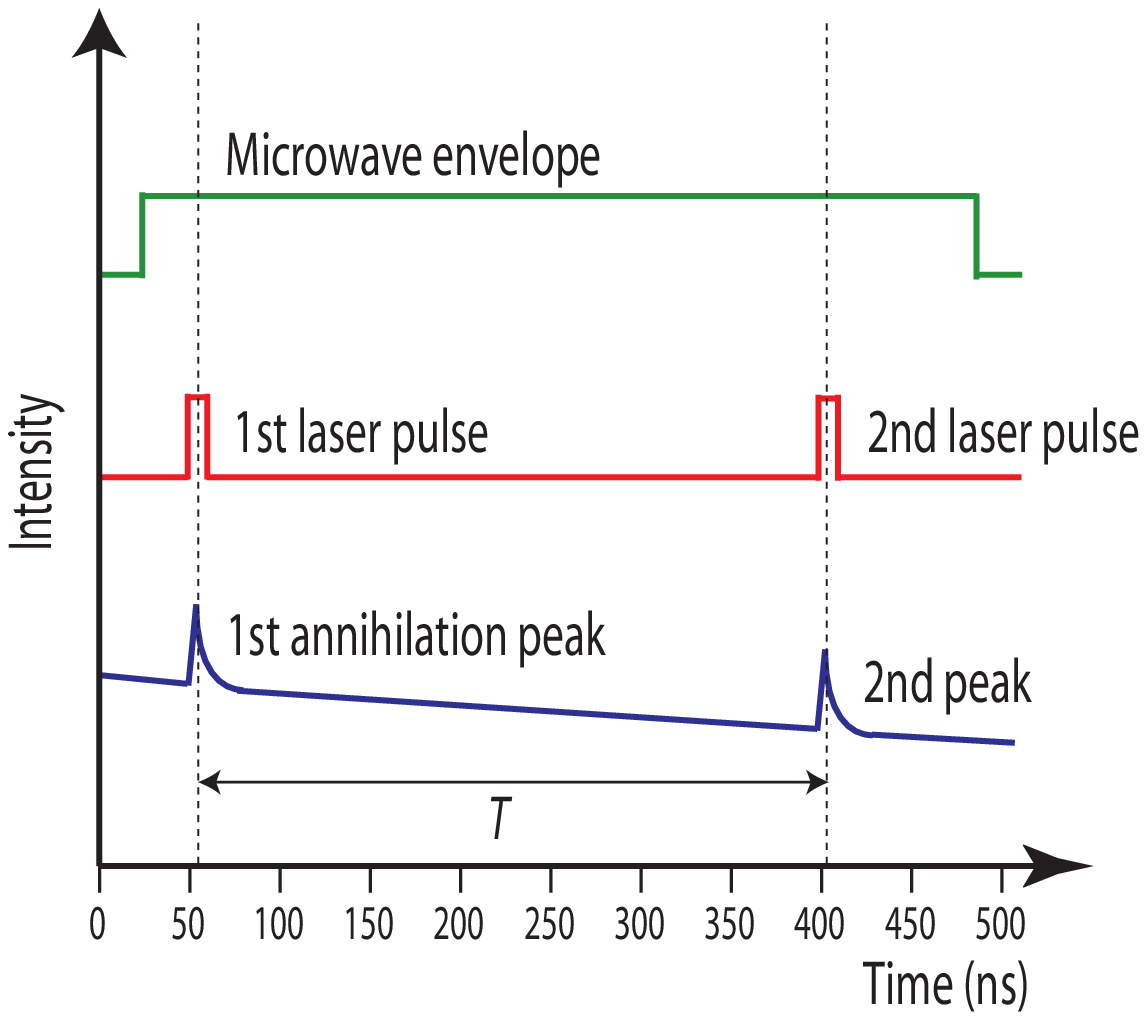}}
\subfloat[]{
\label{fig:subfig:target}
\includegraphics[scale=0.5]{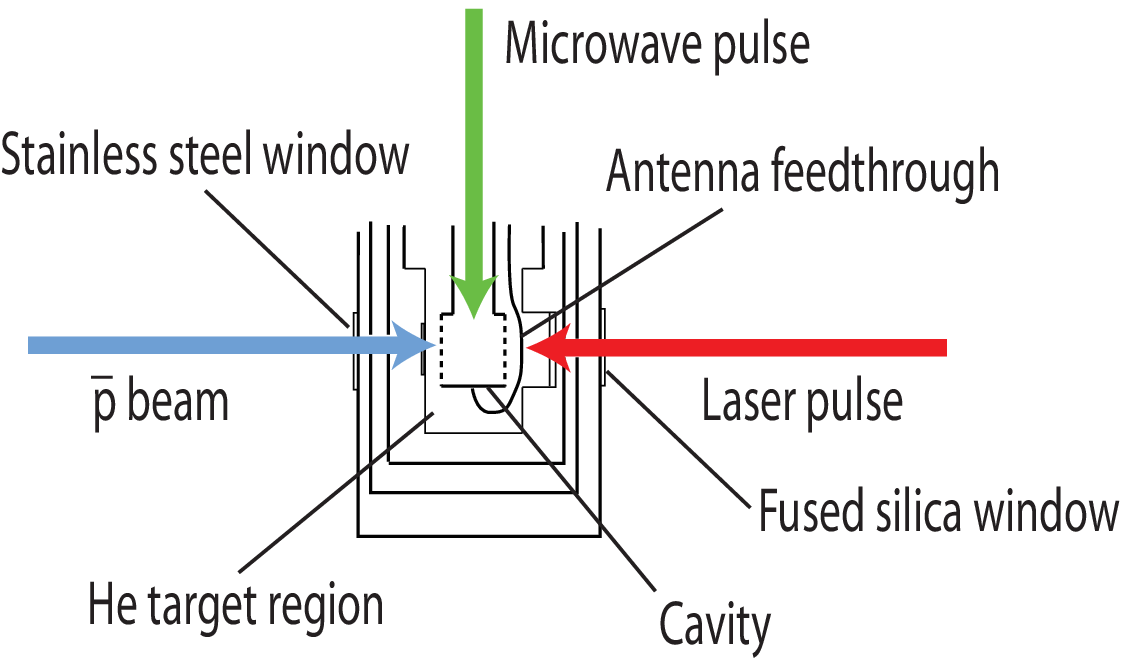}}
\end{center}
\caption{(a)Schematic to show timing between the microwave envelope, the two laser pulses and the resulting annihilation peaks.  The microwave envelope starts before the first laser and ends after the second.  However before firing the first laser there is no population asymmetry and thus no observable transition can occur.  After firing the second laser no further measurements are made so the effective microwave pulse length is equal to the delay between the two lasers $T$. (b) Schematic of the target region.}
\label{fig:ADATS}
\end{figure}

The laser spectroscopy experiments employ a technique by which the annihilation decay products are detected~\cite{Yamazaki:02}.  A sharp prompt peak is first observed, where the majority of states annihilate within picoseconds of formation, then an exponential tail, where the metastable states cascade more slowly towards the nucleus.  This tail constitutes the background.

A narrow band pulsed laser is scanned over the region of an expected transition between a radiative decay dominated parent state and an Auger decay dominated daughter.  Because the daughter state is relatively short lived ($\sim$~10~ns), resonance is indicated by a sharp peak against the background at the time of the pulse.  The ratio of the peak area to this background (peak-to-total) indicates the size of the population transferred.

The HF splitting measurement method is illustrated in Fig.~\ref{fig:subfig:timing} where the first laser pulse remained fixed to the $f^+$ transition between the (37,~35) and (38,~34) states, creating a population asymmetry.  A microwave pulse, if on resonance with either $\nu^{+}_{\rm{HF}}$ or $\nu^{-}_{\rm{HF}}$, then transferred the population from $J^{-+}$ or $J^{--}$ to refill the $F^+$ state.  A second laser pulse was tuned to the same $f^+$ transition and fired with a delay $T$~=~200~-~500~ns from the first, which measured the population transfer.  Plotting the peak-to-total of the second laser induced annihilation peak as a function of microwave frequency yields the two HF transitions as distinct peaks~\cite{Pask:2008}.

\begin{figure}[t]
\begin{center}
\includegraphics[scale=0.35]{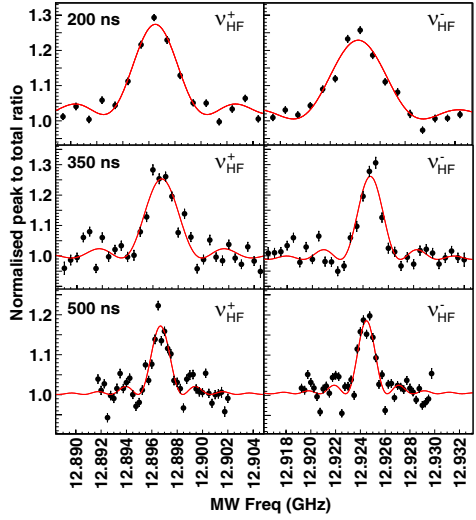}
\end{center}
\caption {Microwave frequency profiles averaged from scans at a common pressure $p$~=~150~mbar and laser delays $T$~=~200~ns, 350~ns and 500~ns.  The broadening is due to the Fourier transform of the rectangular microwave pulse of length $T$.}
\label{fig:All}
\end{figure}

\begin{figure}[!tbh]
\begin{center}
\subfloat[]{
\label{fig:subfig:density_points}
\includegraphics[scale=0.3]{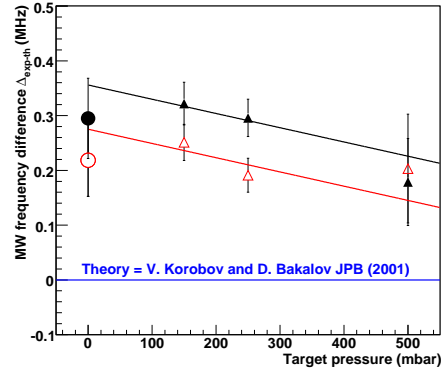}}
\subfloat[]{
\label{fig:subfig:power_points}
\includegraphics[scale=0.3]{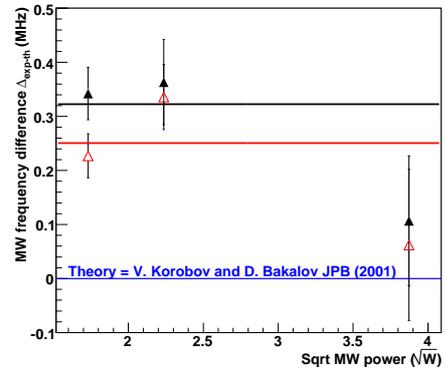}}
\end{center}
\caption {The difference $\Delta_{\rm{exp-th}}$~=~$\nu_{\rm{exp}} \, - \, \nu_{\rm{th}}$ for $\nu^{\pm}_{\rm{HF}}$ between experiment and the closest theory~=~0~MHz~\cite{Korobov:01}.  The second theory \cite{Kino:03APAC} is another 300~kHz less.  The experimental value of $\nu^+_{\rm{HF}}$ is shown as a solid triangle ($\blacktriangle$) and  $\nu^-_{\rm{HF}}$ as an empty triangle (\textcolor{red}{$\vartriangle$}).  The estimated theoretical error is 1.3~MHz~\cite{Bakalov:07} and therefore too large to be shown on the scale of these graphs. (a) Pressure dependence, where the point at 250~mbar is the average of two power dependent measurements from~\cite{Pask:2008}.  The points represented by the solid circle ({\LARGE $\bullet$}) and empty circle (\textcolor{red}{{\LARGE$\circ$}}) shown at $p$~=~0~mbar are $\overline{\nu}^+_{\rm{HF}}$ and $\overline{\nu}^-_{\rm{HF}}$ respectively. (b) Power dependence measured at constant pressure $p$~=~150~mbar the average of which constitutes the point at $p$~=~150~mbar in (a).}
\label{fig:Points}
\end{figure}

The experiment was carried out at CERN's Antiproton Decelerator (AD) which provided a pulsed beam of (1~-~3)~$\times$~$10^{7}$ antiprotons with length $\sim$~200~ns and energy 5.3~MeV.  Every 90~-~120~s, such a pulse was stopped in a helium gas target at a temperature of 6.1~K and pressures $p$~=~150~-~500~mbar (number density 1.7-6.2~$\times$~10$^{20}$~cm$^{-3}$).  The antiproton annihilation products passed through one of two Lucite plates either side of the target where their Cherenkov photons were detected by photomultipliers (PMT)~\cite{CherHori:03} and the signal was displayed on a digital oscilloscope.  The PMTs were gated off for the $\pbar$ arrival so that only the 3\% metastable tail was recorded.

\begin{table*}[!htb]
\begin{center}
\caption{A list of the experimental data including parameters pressure $p$, laser delay $T$, microwave power $P$, number of shots from the AD  $\pbar$ shots, HF transition frequencies $\nu^{\pm}_{\rm{HF}}$, and peak widths $\Gamma_{\pm}$.}\label{tab:List}
\begin{tabular}{l | l | l | l | l | l | l | l}
\hline
\hline
$p$ (mbar) & $T$ (ns) & $P$ (W) & $\pbar$ shots & $\nu^{+}_{\rm{HF}}$ (GHz) & $\Gamma_+$ (MHz) & $\nu^{-}_{\rm{HF}}$ (GHz)& $\Gamma_-$ (MHz)\\

\hline
150 & 200 & 15 & 1070 & 12.896 45(12)  & 3.84(8) & 12.924 30(14)  & 4.70(3)\\
150 & 350 &  5 & 1028 & 12.896 709(78) & 2.76(8) & 12.924 579(59) & 2.21(4)\\
150 & 500 &  3 & 2236 & 12.896 688(48) & 1.68(6) & 12.924 470(41) & 1.48(4)\\
250~\cite{Pask:2008} & 350 &  5 & 2938 & 12.896 651(35) & 2.24(2) & 12.924 431(35) & 2.41(3)\\
250~\cite{Pask:2008} & 500 &  3 & 230 & 12.896 53(12)  & 2.55(7) & 12.924 446(65) & 1.65(5)\\
500 & 350 &  5 & 1844 & 12.896 525(80) & 2.06(5) & 12.924 446(99) & 2.01(7)\\
\hline
\hline
\end{tabular}
\end{center}
\end{table*}

The pulse-amplified continuous wave (CW) laser system~\cite{Hori:06,Hori:09p} was constructed by splitting a CW laser of wavelength $\lambda$~$\sim$~726.1~nm into two seed beams.  These were each pulse amplified by a NG:Yag laser and three Bethune dye cells, the second delayed by a time $T$ after the first.  The pump beams were stretched so that the pulse lengths of the two lasers were 18~ns and 13~ns~\cite{Pask:2008}, slightly longer than the Auger lifetime of the daughter state, to achieve a high depopulation efficiency.

The microwave apparatus was similar to that described in Sakaguchi \emph{et al.}~\cite{Sakaguchi:04} and a schematic is displayed in Fig.~\ref{fig:subfig:target}.  The microwave pulse was synthesised by a vector network analyser (Anritsu 37225B) referenced to a 10~MHz satellite signal (HP~58503B) and amplified by a travelling wave tube amplifier (TMD PTC6358).  A waveguide carried the pulse to a custom made stainless steel cylindrical cavity, with central frequency $\sim$~12.91~GHz, which provided the desired shape for the field (TM$_{110}$ mode) at the target.  Steel meshes (92\% transparency) covered both ends of the cavity so that antiprotions and the two laser beams could enter the target from opposite directions.  The cavity was overcoupled to the waveguide to achive a broad frequency range $\sim$~100~MHz.  A mu-metal shell surrounded the target region to protect from external magnetic fields.  Indeed the field measured in three dimensions within the target was $B$~$<$~0.03~G.

Previously, different choke positions of a triple-stub-tuner were used to match the impedance of the waveguide to that of the cavity for a range of frequencies~\cite{Sakaguchi:04}.  This time, a constant microwave power $P$ was produced at the target by firing a predetermined signal strength down the unmatched waveguide.  Most of the signal was reflected and dumped to a 50~$\mathrm{\Omega}$ terminator by a three-way circulator.  This removed standing waves from the system and allowed the relatively small amount of power absorbed by the cavity to be controlled to within 1~dB over the frequency range.  The power was monitored by an undercoupled pickup antenna situated opposite the waveguide.

Table \ref{tab:List} shows a summary of all data measured for this experiment.  The line shape was determined in Pask \emph{et al.}~\cite{Pask:2008} to be 
\begin{displaymath}\label{eq:linewidth}
X(\omega) = \frac{|2b|^2}{|2b|^2+(\omega_0-\omega)^2}\sin^2 \bigg\{ \frac{1}{2} \Big[|2b|^2+(\omega_0-\omega)^2 \Big]^{\frac{1}{2}}T \bigg\}.
\end{displaymath}
\noindent Data measured with the same $p$ and $T$ were fitted simultaneously with common parameters for height, width and central frequency.  Two  data sets were systematically examined: 1) Microwave power dependence, and 2) Pressure dependence

1) The ac Stark effect shifts the E1 transitions by less than one part in  $10^9$~\cite{Hori:06}.  Its equivalent, the ac Zeeman shift of the  M1 transitions is far weaker and therefore far too small to be resolved.  A power dependence measurement was nevertheless examined for a complete understanding of the systematics.  At a constant pressure $p$~=~150~mbar, resonance profiles were measured with various laser delays $T$~=~200~ns, 350~ns and 500~ns at microwave powers carefully chosen to achieve a $\pi$-pulse~\cite{Pask:2008p}, $P$~=~15~W, 5~W and 3~W, respectively.  For illustrative purposes the average of these scans is shown in Fig.~\ref{fig:All} where the the dominating broadening effect was due to the Fourier transform of the rectangular microwave pulse of length $T$~\cite{Pask:2009p}.  Figure~\ref{fig:subfig:power_points} shows that no power dependent trend was observed, therefore all data measured at common target densities could be averaged.

2) In 2006 Korenman predicted a collisional shift of $\Delta$~$\equiv$~d$\nu/$d$p$~=~0.3~kHz/mbar~\cite{Korenman:2006} and, for the first time, a resolution has been achieved to examine this.  High statistic microwave resonance profiles were measured at $p$~=~150~mbar and 500~mbar.  Previous measurements had been made at $p$~=~250~mbar~\cite{Pask:2008}.  

\begin{figure}[!tbh]
\begin{center}
\includegraphics[scale=0.3]{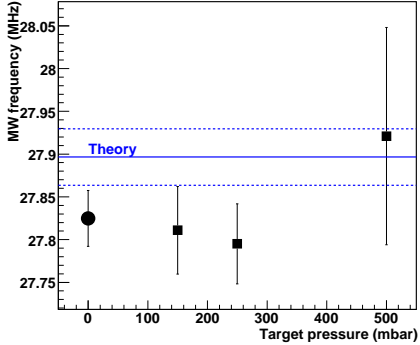}
\end{center}
\caption {The difference for $\Delta \nu_{\rm{HF}}$ between experiment and theory~\cite{Korobov:01,Kino:03APAC} as a function of target pressure.  The experimental values are shown as squares ($\blacksquare$).  The point shown at $x$~=~0 and represented by a circle ({\LARGE $\bullet$}) represents the average of the total data.}
\label{fig:Diff}
\end{figure}

The results are displayed in Fig.~\ref{fig:subfig:density_points} and clearly show that, if any such density dependence existed, the trend may have the same magnitude but opposite sign.  Extrapolating the points to zero density can be performed with lines of average gradient $\Delta$~=~$-$0.26~$\pm$~0.2~kHz/mbar.  Such a large error neither confirms nor precludes conclusively the existence of a shift.  The two fits in Fig.~\ref{fig:subfig:density_points} are shown with the same gradient because no density shift was observed for the difference between the transitions $\Delta \nu_{\rm{HF}}$~=~$\nu^{-}_{\rm{HF}}$~$-$~$\nu^{+}_{\rm{HF}}$, see Fig.~\ref{fig:Diff}, in accordance with predictions.

A more recent calculation~\cite{Korenman:2009p}, based on measurements of the collisional broadening~\cite{Pask:2009p}, predicts $\Delta$~=~$-$0.048~kHz/mbar which is less than the experimental precision.  As a negative slope contradicts the predictions, the average of each measurement was taken but the error used was that of the extrapolation to zero density from the fit.

Other systematic effects that influence the measurement include external magnetic fields, precision of the microwave frequency source, shot-to-shot microwave power fluctuations and variances in the laser position and fluence from day to day.  However these effects have been determined to be far smaller than the shot-to-shot fluctuations of the antiproton beam.  Data was measured over a long period to reduce these drift effects and variations in the $\pbar$ intensity have been reduced by normalising the second laser induced annihilation peak with the first (proportional to the number of antiprotons captured).  Despite these considerations the reduced chi-squared $\chi_{\rm{red}}$ of the fit was $\chi_{\rm{red}} \, \sim \, 3$.  To adjust for this the error bars were inflated  by $\sqrt{\chi_{\rm{red}}} \, \sim \, 1.7$.

Bakalov calculated that a broadening  of $\nu^{\pm}_{\rm{HF}}$ due to an external magnetic field occurs at a rate of $\Gamma_{\pm}$~$\sim$~5.6~MHz/G~\cite{Bakalov:2009p}. The similarity between the  Fourier transform of the microwave pulse and the spectral line widths~\cite{Pask:2009p} confirms that the target region was well shielded during the experiment.  Due to referencing to a 10~MHz GPS receiver, the precision of the frequency source is several orders of magnitude less than the resolution of this experiment.  Thefore the statistical errors are much greater than the systematic.

\begin{table*}[!bth]
\begin{center}
\caption{Experimental data compared with three-body QED predictions, where $\nu^{\pm}_{\rm{HF}}$ are the HF transition frequencies and $\Delta \nu_{\rm{HF}}$ is the difference between  $\nu^{-}_{\rm{HF}}$ and $\nu^{+}_{\rm{HF}}$.  The quoted theoretical errors have been estimated by Bakalov and Widmann~\cite{Bakalov:07}.}\label{tab:Results}
\begin{tabular}{l | l | l | l}
\hline
\hline
 & $\nu^{+}_{\rm{HF}}$ (GHz) & $\nu^{-}_{\rm{HF}}$ (GHz)& $\Delta \nu_{\rm{HF}}$ (MHz)\\

\hline
This work & 12.896 641(63) & 12.924 461(63) & 27.825(33) \\
2002~\cite{Widmann:02} & 12.895 96(34) & 12.924 67(29) & 28.71(44) \\
\hline
Korobov~\cite{Korobov:01} & 12.896 3(13) & 12.924 2(13) & 27.896(33) \\
Kino~\cite{Kino:03APAC} & 12.896 0(13) & 12.923 9(13) & 27.889(33) \\
\hline
\hline
\end{tabular}
\end{center}
\end{table*}

The individual transition frequencies have a negligible dependence on $\mu^{\pbar}_{s}$.  However $\Delta \nu_{\rm{HF}}$ is directly proportional to this value.  The predicted density shift for $\Delta \nu_{\rm{HF}}$ is far smaller, $\Delta$~=~0.003~Hz/mbar~\cite{Korenman:2009p} than the precision of this experiment.  If this is the case, the total splitting can be calculated from the difference between each pair of transitions that were measured at common densities $\Delta \nu_{\rm{HF}}$~=~$\sum^N_i (\nu^{-}_{\rm{HF}_i}$~$-$~$\nu^{+}_{\rm{HF}_i})/N$, rather than the difference between the sum of each transition measurement $\Delta \nu_{\rm{HF}}$~=~$(\sum^N_i\nu^{-}_{\rm{HF}_i}$~$-$~$\sum^N_i\nu^{+}_{\rm{HF}_i})/N$, where $i$ is the index of a measurement and $N$~=~3 is the total number of density dependent measurements.  Figure~\ref{fig:Diff} displays $\Delta \nu_{HF}$ as a function of target pressure compared to the two most recent theories.

Fitting a first order polynomial, results in a gradient almost half that of its associated error, $\Delta$~=~0.24~$\pm$~0.37 kHz/mbar, so the above holds and the data can be averaged to obtain a final value of $\Delta \nu_{\rm{HF}}$. Table~\ref{tab:Results} presents the data for the recent and previous experiments compared to the two most up to date theories.

This work demonstrates the completion of a systematic experimental study on the HF splitting of the (37,~35) state of $\pbHe$.  The experimental error of $\nu^{\pm}_{\rm{HF}}$ has been reduced by a factor 20 less than that of the theoretical calculations and, although $\Delta_{\rm{exp-th}}$~=~300~-~600~kHz, it is well within the estimated theoretical error 1.3~MHz.  The experimental precision for $\Delta \nu_{\rm{HF}}$ has reached that of theory and has been improved by more than a factor of 10 over the first measurement~\cite{Widmann:02}. There is a two sigma agreement between theory and experiment.

The sensitivity $S$ of $\Delta \nu_{\rm{HF}}$ on $\mu^{\pbar}_{s}$ for the (37,~35) state is  $S$~$\equiv$~d$E$/d$\mu^{\pbar}_{s}$~=~10.1~MHz/$\mu_{N}$~\cite{Bakalov:07}, where $\mu_{N}$ is the nuclear magneton. Thus the magnetic moment can be determined to be:

\begin{equation}
\mu^{\pbar}_{s} \, = \, -2.7862 (83) \mu_{N},
\end{equation}

\noindent where the uncertainty has been calculated by adding $\Delta_{\rm{exp-th}}$ with the errors of theory and experiment in quadrature, resulting in a one sigma error, slightly less than the value determined by Kreissl \emph{et~al.}~\cite{Kreissl:88}, while the deviation from the magnitude of the proton spin magnetic moment, $\mu^p_s \, = \, 2.792847351(28)$, is similar but opposite in sign.

The absolute values for the magnetic moments of the proton and antiproton are in agreement within

\begin{equation}
\frac{\mu^p_s \, - \mid \mu^{\pbar}_{s} \mid}{\mu^p_s} \, = \, (2.4 \, \pm \, 2.9) \, \times \, 10^{-3}.
\end{equation}

The limit of experimental precision has been reached for the (37,~35) state.  The study of other states with larger $\Delta \nu_{\rm{HF}}$ could potentially increase the precision but the system cannot generally be improved due mainly to uncontrollable fluctuations of the AD beam.  Preparations are underway to measure the HF splitting of $\pbar ^3 \rm{He}^+$ which, because of the additional helion spin, provides a more thorough test of the theory but yields no further information on the magnetic moment.  The theorists are currently working on an $\alpha^6$ calculation but a significant change in $\Delta \nu_{\rm{HF}}$ is not expected.\\

The authors would like to acknowledge V.~Korobov (Joint Institute for Nuclear Research, Russia), D.~Bakalov (INRNE, Bulgaria) for many helpful discussions.  We thank two undergraduate students: P.~Somkuti and K.~Umlaub who contributed to this project.  We are also grateful to the AD operators for providing the antiproton beam.  This work was supported by Monbukagakusho (grant no. 15002005), by the Hungarian National Research Foundation (NK67974 and K72172), the EURYI Award of the European Science Foundation and the Deutsche Forschungsgemeinschaft (DFG), the Munich-Centre for Advanced Photonics (MAP) Cluster of DFG and the Austrian Federal Ministry of Science and Research.



\bibliography{ps205,hbar,EBW-new,DH-new,RSH-new}

\end{document}